\documentstyle[11pt,oldlfont]{article}

\def\R{\mbox{\boldmath$R$}}

\def\R{\rlap{\rm I}\mkern3mu{\rm R}}
\def\Z{\rlap{\sf Z}\mkern3mu{\sf Z}}
\begin{document}
\thispagestyle{empty}

\begin{flushright} 
                    LPT-ENS 02/42 \\
                     \end{flushright}

\vspace*{0.5cm}

\begin{center}{\LARGE {U-opportunities : why is ten equal to ten ?}}

\vskip1cm

Bernard Julia$^a$  

\vskip0.5cm

$^a$Laboratoire de Physique Th{\'e}orique CNRS-ENS\\ 
24 rue Lhomond, F-75231 Paris Cedex 05, France\footnote{UMR 8549 du
CNRS et de l'{\'E}cole Normale Sup{\'e}rieure.
}\\
\vskip0.5cm

\begin{minipage}{12cm}\footnotesize

{\bf ABSTRACT}

\bigskip

It seems to me at this time that two recent developments may permit fast progress on our way to understand 
the  symmetry structure of toroidally (compactified and) reduced M-theory. The 
first indication of a (possibly) thin spot in the wall
that prevents us from deriving a priori the U-duality symmetries of these models is to be found in the analysis of  the hyperbolic billiards that control  the chaotic time evolution of (quasi)homogeneous anisotropic String, 
Supergravity or Einstein cosmologies near a spacelike singularity. What happens is that  U-duality symmetry controls chaos via negative constant curvature. On the other hand it was noticed in 1982 that  (symmetrizable) ''hyperbolic'' Kac-Moody algebras have maximal rank ten, exactly like superstring models and that two of these four rank ten algebras matched physical theories.  My second reason for optimism actually predates also the previous breakthrough, it was the discovery in 1998 of surprising superalgebras extending U-dualities to all (p+1)-forms (associated to p-branes). They have a super-nonlinear sigma model structure
similar to the symmetric space structure  associated to 0-forms and they obey a universal self-duality field equation. As the set of forms is doubled to implement electric-magnetic duality, they obey a set of first order equations. 
More remains to be discovered but the beauty and subtlety of the structure 
cannot be a random emergence from chaos. In fact we shall explain how a third 
maximal rank hyperbolic algebra $BE_{10}$ controls heterotic cosmological chaos
and how as predicted Einstein's General Relativity fits into the general 
picture. 
\bigskip
\end{minipage}
\end{center}
\newpage

\section{Classifications}

It is well known to conformal field theorists but a much more general and venerable fact that positive definite symmetric matrices with integer entries tend to appear in many classification problems. More precisely the ADE
 philosophy is to list such occurrences and try to relate them to each other. 
Let us be a bit more specific, this set of  problems corresponds to the 
emergence in various contexts of symmetric
matrices with diagonal entries equal to 2 and negative integral off-diagonal 
entries. The prototype of a successful correspondence is the work of 
Brieskorn (with some help from Grothendieck) realising the simple complex
Kleinian singularities related to discrete subgroups of SU(2) 
as  rational singularities in the set of unipotent elements of the corresponding complex Lie group of type ADE.
The singularity is at the conjugacy class of subregular elements in other words the elements whose centraliser
  has  2 more dimensions than regular ones namely $r+2$ instead of  $r$,
 the group rank.  
The resolution of the singularity introduces exceptional divisors whose intersection matrix is the opposite of the group Cartan matrix, there the ADE matrices arise in two different disguises but there is a relation to the Lie group in both cases. 
   
Halfway (in time) between the Brieskorn results and the classification of 
modular invariant partition functions by Cappelli et al. $N=8$ supergravity 
was constructed in 4 dimensions as well as its toroidal decompactification 
family up to 11 spacetime dimensions. It was quickly remarked to me by  Y. 
Manin that the $E_r$ internal symmetry groups appearing upon compactification 
on a r-torus suggested a role for (regular) del Pezzo surfaces, these are 
variants of $CP_2$ or $CP_1 \times CP_1$ which admit a canonical projective 
embedding . However the other pure supergravities in 4 dimensions with fewer 
supersymmetries also belonged to families which together formed a magic 
triangle of theories. They led  also to ADE groups yet not all 
in split form, this meant that real geometry was to be tackled rather than 
the simpler complex algebraic one.

It also rapidly became clear if not rigorously established that the $E_r$ family included the infinite dimensional 
$E_9 \equiv E_8^{(1)}$ in 2 dimensions ie. after compactification on a 9-torus. Our partial understanding of these duality groups includes their $A_{r-1} \times \R$  subgroups related to the r ignorable coordinates.
The infinite dimensional hidden symmetry is related to the existence of a Lax pair in 2 residual dimensions
 leading to a quasi integrable situation yet it was known that chaos remained  prevalent in some particular flows.

In 1982 I noticed that a naive extrapolation to 1 dimension would suggest a  role for $E_{10}$ as one would expect its subalgebra  $A_9$ to appear there, however the implementation was problematic. Nevertheless a similar analysis of 
so-called type I supergravity in other words pure (without matter multiplet) supergravity in 10 dimensions led to the suggestion of 
a corresponding role for the other hyperbolic  Kac-Moody algebra 
$DE_{10}$ the "overextended" $D_8$ in split form (namely $SO(8,8)$ affinized 
with one more $SL(2,\R)$ generating subgroup right next to the affinizing   
$SL(2,\R)$ in the Dynkin diagram as implied by the $A_{r-1}$ argument). Now 
according to Bourbaki (actually Chein) $E_{10}$ and $DE_{10}$ are exactly the 
two simply laced hyperbolic Kac-Moody algebras of maximal rank. There are only 
two more (non simply laced) hyperbolic Kac-Moody algebras of maximal rank: 
$BE_{10}$ (see section 3) and $CE_{10}$ (to be seen). 

Let us recall
that hyperbolicity here means that the Dynkin diagram becomes a product of 
finite or affine Dynkin diagrams after removal of any node (this gives nice 
arithmetic properties as well). The rank ten here is directly 
related to the rank eight of $E_8$, the 
largest exceptional simple Lie group which itself  comes about also from some 
subtle classification analysis. On the other hand 
ten is the critical dimension of superstring models and as such it is related to quantum conformal  invariance. How could the two derivations be related? They must be as the rank of the Lie group is closely related to the dimension of the compactification torus! 

Still a puzzling fact emerged from the subsequent construction of heterotic 
strings namely the possibility of duality groups of rank higher than 8 for 
instance  16 in dimension 3 ($SO(8,24)$) corresponding  putatively to rank 18 
in one dimension.
This paradoxical apparent violation of the bound ten on the rank will be clarified in  section 3.  Let us also remark that we are talking of superstrings and (bosonic) Kac-Moody algebras, hence fermionic structures should emerge in an a priori  bosonic context.
The upper limit 26 does not arise so simply  yet, although it can be argued to 
be the sum of ten and sixteen the latter being the rank of the two even 
euclidean unimodular lattices dictated again by quantum anomaly considerations.

Let us add some examples of ADE objects. The basic one is the set of integral  
positive definite matrices occuring as Cartan matrices of simple simply laced 
complex Lie groups. The positive definiteness, resp positive semi-definiteness,
resp hyperbolicity guarantees a relatively simple classification, for instance in the positive definite case (that of finite dimensional simply laced Lie algebras) there is a finite number of  objects for each rank; on the other hand the three cases of   $E_6, E_7, E_8$  look exceptional in this context. In the semi-definite case also called
affine Kac-Moody situation very much the same is true; in the hyperbolic case 
however as we have seen the rank is bounded and there are finitely many 
instances of a given rank  except when it is equal to two. 

The finite dimensional (irreducible) Coxeter groups generated by reflections
are closely related objects, their list
encompasses that of  the  Weyl groups of the simple Lie groups but one gets 3
extra Coxeter diagrams with  rotation angle $2\pi/5$ and an infinite family of 
dihedral groups of rotation angles $2\pi/k$ for non cristallographic k's 
integers at least 7. These are all nonsimply laced cases.
It is important to note that non simply laced Lie groups have a symmetrisable 
Cartan matrix: a basic assumption for most of Kac-Moody theory and Borcherds 
algebras. The reflections preserve a symmetric form that is a symmetrisation of the non symmetric Cartan  matrix.

By definition Coxeter groups admit a finite presentation by involutions $S_i 
\, , \, i=1,\dots,r$ satisfying 
\begin{equation}
( S_i S_j)^{m_{ij}} =  I \, , \, i\neq j.
\end{equation}
In the simply laced case we consider only exponents $ m_{ij} =2$ for commuting involutions or 3 for dihedral subgroups of order 6.
The matrix is encoded by a Dynkin diagram with r vertices and simple bonds for 
exponent 3; it turns out that no loop is allowed, that at most three legs occur 
and finally that the sum $1/m+1/n+1/p$ of the inverses of the  number of 
vertices (including the potentially trivalent vertex) on each leg must be strictly larger than one. One recognises two infinite families $A_k, D_k$ and three exceptions $E_6, E_7, E_8$ with numbers of vertices respectively 
$(m,n,p)=(2,3,3), (2,3,4), (2,3,5)$, let us  notice that $E_5\equiv D_5$ corresponds to $(2,3,2)$ and  $E_4 \equiv A_4$  
 to $(2,3,1)$. $E_3\equiv A_2 \times A_1$ which is semi-simple but not simple is the next group in the $E_r$ family.   

The list of affine Kac-Moody algebras is very closely related to the list of finite dimensional simple Lie algebras.
It permits one loop for the  $A_k^{(1)}$ Dynkin diagrams, sums of inverse 
numbers of vertices at 
one three-valent vertex equal to one,  two three valent vertices for 
$D_k^{(1)} \,  k\geq 5$ resp. one 4-valent vertex for $D_4^{(1)}$. The list of 
hyperbolic diagrams can be found in \cite{Og,Sa}. Their defining property given above guarantees the Lorentzian signature of the invariant bilinear form on the Cartan subalgebra. The hyperbolic algebras one meets in supergravity theories
are overextensions of finite Lie algebras, the construction was discussed in 
\cite{J82, F83}. Not all overextensions are hyperbolic but all hyperbolic 
algebras of rank at least 7 are overextensions (called superaffine in 
\cite{Og}).  The derivation of the signature is straightforward, 
if $A$ is a kxk affine Cartan matrix it has  null determinant and signature 
$(+^{k-1}, 0)$ , the (k+1)st line and column contain a 2 at their intersection and a
$-1$ at the affine column or row,  the corresponding quadratic form after 
completing the square has manifest Lorentzian signature. 
Another 
characteristic property we shall use in section 3 is the fact that the Weyl 
chamber on the unit hyperboloid has finite volume (it may actually be  non 
compact,  for instance when the rank is strictly larger than 5 \cite{Bo,Vi}). 

Let us now discuss in more detail the case of overextended $A_{k-1}$. 
The affine $A_{k-1}^{(1)}$ has a circular Dynkin diagram, its 
(over)extension has just one line and one extra vertex attached
to it, now the criterion of hyperbolicity prevents overextended $A_8 $ to be hyperbolic, overextended $A_7$ is the
last hyperbolic  $HA_9$ in the family, and this can be viewed again
as a consequence of the fact that there are only three Lie groups of type E. So one may say that the exceptionality of the E family is related to the bound on the rank of  the hyperbolic $HA_{k+1}$. 
We shall see in section 3 the dramatic difference between pure gravity dynamics at a singularity in ten or eleven dimensions as a consequence of this algebraic fact. 

Let us recall that $E_{10}$ and $HD_{10}$ were identified in \cite{J82}, the Weyl chamber of $HB_{10}$
(in other words overextended $B_8$) was suggestively recognised by \cite{DH}  as controlling the  classical chaos of heterotic string theory yet following Narain and Sen one expects  the U duality group to have rank 
16 in  3 dimensions  and not 8. The answer lies in the simple observation that the classical action is a real functional
and the real equations are invariant under a real Lie group, the precise real 
form of which is critically important. But as an 
expert (M. Reid) puts it, real algebraic geometry is $2^N$ times more 
difficult than complex algebraic geometry with N large.  

\section{Real forms of Lie algebras}

We are familiar with the classification of complex simple Lie algebras as a monument of group theory. Its     relative simplicity is permitted by the algebraic closure of the field of complex numbers, indeed the main tool is the simultaneous diagonalisation of commuting Cartan generators (observables) and the analysis of the root spectrum (quantum numbers). Up to central elements (non simple-connectedness) and  up to isomorphism there is a unique compact form of the associated Lie group. 
The theory of non compact forms and their representation theory is much richer and even the split (also called maximally non-compact forms) have 
complicated representations. The existence of the split form follows from the observation that the structure constants of a complex Lie algebra can be taken to be integers in the appropriate basis. The restriction of the field of coefficients from 
the complex to the real numbers or even the natural integers is possible, the 
choice of real ''angles of rotations"  in the Cartan-Chevalley basis defines 
the split form. For $SL(2,C)$ the split form is $SL(2,\R)$ whereas the compact 
form is $SU(2)$ and the arithmetic form $SL(2,\Z)$.

Over the complex numbers or in the compact case the Cartan subalgebras are all conjugate to each other, not so for other real forms $G$ even the split one, still there is a finite collection of inequivalent ones. The classification of real forms of simple Lie groups has been given by E. Cartan  together with the classification of maximally symmetric spaces. Since then two strategies have been 
applied. Either one selects a maximally non-compact Cartan torus in $G$ and 
diagonalises as many observables as possible over the reals, there appears a 
maximal split subalgebra $S$ (\cite{BT} p.116) inside our noncompact real form 
$G$ and the roots project onto roots or twice the roots of  $S$, this is the 
Tits-Satake theory with 
bicoloured diagrams as developped for instance in \cite{Ar}. 
One key result is that the roots restricted to the noncompact Cartan generators 
form a not necessarily reduced root system. In particular this implies that 
after choosing the compact real form that contains the Cartan  (compact) 
generators resp. their multiples by the imaginary unit i (for the non-compact 
ones), all "imaginary roots"
ie. those that vanish on the (maximal) set of non-compact Cartan generators must be associated to compact  eigengenerators.

The other frequent choice of maximal torus leaves more freedom, given a 
non-compact real form one chooses a maximally compact Cartan 
subalgebra. Vogan  introduced other bicoloured diagrams for that situation.
 For instance the case of fully compact tori arises sometimes for non-compact 
Lie algebras and is very interesting. Now the arbitrariness with the Vogan  
strategy  stems from the fact that the choice of simple roots is not unique 
there. The bicoloured diagrams may be chosen to have all or all but one compact vertices.  Clearly this choice permits an easier identification of compact subgroups whereas the Tits-Satake strategy is best for split subgroups. It is natural
 to expect a dual result  to the previous one namely that now 
no root can vanish 
on the (maximally) compact part of the Cartan  torus, in other words one departs
maximally    from the complex root analysis. We refer to \cite{Kn} for an introduction to the Borel-de Siebenthal-Murakami-Vogan theory.

The real rank  $l$ of a simple real Lie algebra of full rank $r$ is the maximal dimension of an abelian subalgebra of diagonalisable (called semi-simple) 
generators of non-compact type (whose Killing norm is positive). 
$0\leq l \leq r$, with $r=l$ in the split case and $l=0$ in the compact case. 
If one starts from the split form the compact form is obtained by  multiplying some generators by $\sqrt{-1}$. For instance given the standard basis $e,f,h$ of $SL(2,\R)$ the compact $SU(2)$
  is generated by $(e-f) $ which is already compact as well as  $i(e+f)$ and $ih$. We see that the diagonalisability of $ih$ has been lost over the real numbers.  

Turning now to the  applications we may  read off the tables of \cite{Ar,He} 
that the restricted root system associated to
 the non-split $E_7(-5)$ of real rank 4 (the U-duality symmetry of N=6
  4d supergravity reduced to three dimensions) is that of its  maximal split 
subalgebra $F_4$ 
(one speaks of Freudenthal-Tits geometry of type $F_4$) and this is a common 
feature  of all Maxwell-Einstein N=2, d=5 supergravities constructed by 
G\" unaydin et al. in 1983 after reduction to three dimensions.
We refer to \cite{He} p.534 for the Tits Satake diagram of type EVI symmetric 
space or real form of $E_7$. On this diagram the white dots denote non compact 
$SL(2,\R)$ subalgebras, three of them building up a compactification
$SL(4,\R)$ symmetry. The black dots refer to compact $SU(2)$'s. 

In the tables one finds also the multiplicity of the restricted roots; for 
instance for the Lorentz group  $SO(1,3)$  of real rank 1
the Tits-Satake diagram is composed 
of two disconnected white dots with one double arrow between them. 
This can be understood as follows, the root analysis over the real numbers goes 
through for the noncompact (white) Cartan generator(s). Upon projection of 
the full root system on its restriction to linear forms over 
the noncompact part of the
Cartan subalgebra this may imply multiple occurrence of the same 
restricted roots, when this happens for two simple roots one joins the corresponding (white) dots by a double arrow.
The real Lie algebra admits a complex structure precisely when all restricted 
roots have multiplicity two. On the other hand black dots project to zero.
When the restricted root system is not reduced, ie for some geometries of type
$B_k$ one must also give the 
multiplicity of the doubles of the restricted roots.

Our present interest in the Tits-Satake analysis stems from the 
observation that for the orthogonal groups of the form
$SO(8,8+p)$ with p between 1 and 16 the real rank is 8 and the geometry is of 
type $SO(8,9)$. Now such groups arise  as U duality groups in three dimensions 
and the sigma model Lagrangians that control chaos in the quasi homogeneous 
situation must lead to the overextension $BE_{10}$ if one is to recover the experimental discovery of \cite{DH}.
 
\section{Chaos controlled by symmetry}

In a remarkable analysis \cite{DH} it was found that near a cosmological 
singularity the chaotic behaviour of essentially one dimensional (homogeneous)
string theories was well approximated by an Anosov flow in the hyperbolic 
billiard defined by the Weyl cell of hyperbolic Kac-Moody algebras. For M-theory
(alias in this approximation 11d SUGRA or type II String theory) $E_{10}$ 
emerged, and for type I SUGRA $DE_{10}$ replaces it. As we mentioned above both
Lie algebras had surfaced as tantalising candidates for hidden symmetries in 
exactly these situations 20 years ago. Symmetry controls chaos which is not 
so surprising when negative curvature non-compact symmetric spaces are the 
arena. What remains more puzzling is the precise sense in which the 2
dimensional reductions of these models can be called integrable and still 
allow in their midst chaotic islands: they are well known to admit Lax pairs...

The chaotic solutions are even older and this tension between order and 
ergodicity is one of the most striking features of these Lagrangian theories.
Another surprise of this work was the emergence of $BE_{10}$'s Weyl cell as the 
billiard  relevant for heterotic and type I cosmological singularities. The 
relevant U-duality group in three dimensions is $SO(8,24)$ and its 
overextension is not hyperbolic. How should the overextension of $SO(8,9)$
take its place. On the other hand it was also remarked in \cite{J82} that 
$BE_{10}$ and $CE_{10}$ were nonsimply laced analogues of the previous two 
maximal rank algebras and ought to appear somewhere, half this prediction has 
been now fulfilled. We shall explain momentarily the reconciliation of 
$SO(8,24)$ and $SO(8,9)$. 

In \cite{J82} it was also predicted that the overextension $HA_3$ of 
$SL(2,\R)$, the Ehlers group of stationary General Relativity, 
was the more conservative candidate for a hidden symmetry but now 
in a well tested theory  and again in the homogeneous situation. It was a very 
powerful experience 
to meet Alex Feingold and Igor Frenkel in Chicago that summer who independently 
and for mathematical reasons were working on \cite{F83} and had developed the 
theory of superaffine algebras (better called hyperaffine maybe) as a handle 
for hyperbolic algebras while I had been concentrating on the $A_{k-1}$ concept
for physical reasons. We learned a lot from each other then and it is a good 
place to express my thanks to the organisers I. Singer, P. Sally, G. Zuckerman, H. Garland and M. Flato for their invitation. Now this experience 
immediately suggested that the corresponding billiard should appear in the 
celebrated Belinsky-Khalatnikov-Lifschitz chaos. This time it appeared 
only in the 
quasi-homogeneous situation ie not quite in the 1-dimensional setting,
yet 
again the three dimensional U-duality controlled chaos by the same mechanism
\cite{DHJN}. Furthermore the earlier observation by the belgian team of the 
absence of BKL chaos for pure gravity beyond 10 dimensions exactly matched the 
above remark about $HA_9$ as the largest hyperbolic overextension of a group 
of A type. It was known that $A_k$ is precisely the U-duality group of the  
three dimensional reduction of pure gravity in k+3 dimensions \cite{CJLP}.
 
Let us now see how the dimensionally reduced action of such a gravity theory
(on a torus by  homogeneity) implies the dynamical mechanism that is 
approximated by a hard walled billiard for time evolution near a singularity.
We shall begin with the case of split U-dualities ie the better studied examples of \cite{CJLP}. As is well known in three dimensions all propagating fields can be dualised to scalars and form a noncompact symmetric space. There is a choice
between two descriptions here. Either as it is the case when one actually 
discovers the system upon dimensional reduction one works in a fixed (or 
partially fixed) gauge and 
one uses coordinates on the symmetric space; for instance by using the Iwasawa 
decomposition one may use affine coordinates on the Borel subgroup $AN$ of the 
noncompact group $G=KAN$ where $K$ is the maximal compact subgroup of $G$.
Or else one may restore a local $K$ gauge invariance and use $G$ valued scalar 
fields, this is the better  way to discover and restore symmetries but the 
formalism developed in \cite{CJLP} and references therein is best suited for 
the fixed gauge approach.

The general action distinguishes the dilaton fields ie the coordinates along the
(fully) non-compact Cartan subalgebra $Lie(A)$ from the other scalar fields 
that correspond to positive root generators. 
It is simply a sum of quadratic kinetic terms for the latter
weighted by appropriate exponentials of their respective roots ie. linear 
forms in the dilatons plus free 
kinetic terms for the dilatons themselves. These exponential factors are 
responsible for the walls of the effective potential of the piecewise
Kasner metric evolving in time \cite{DH, DHJN} after suitable overextension, in particular one must include the so-called symmetry wall. 

Let us now recall the Iwasawa decomposition in the general case. 
It can be studied in \cite{He} for 
instance. We are now considering a real Lie algebra with a maximally 
non-compact Cartan subalgebra whose noncompact part we denote by  $a$. 
The maximal compact subgroup has a higher dimension than in the split case 
and the coset space has lower dimension. The decomposition reads now 
\begin{equation}
Lie(G)=Lie(K)+a+n
\end{equation}
where $n$ is the nilpotent subalgebra of positive restricted root vectors.

Clearly in the non-split case the Cartan subalgebra is to be replaced by its non compact part
and the nilpotent subalgebra of positive root vectors by that of positive
 restricted root vectors. Again the restricted roots form a possibly nonreduced
 root system.
This brings two possible complications: firstly the multiplicities but they do 
not change the walls of the billiards and hence are irrelevant but also the nonreduced roots
for some $SO(2k+1)$ restricted root systems and geometries.

Applications will appear soon, firstly in \cite{HJK} we shall examine 
the replacement of $SO(8,24)$ by 
$SO(8,9)$; this paper contains  also anomaly free string realisations of the 
$DE_{10}$ 
theory as well as that of a theory with U-duality $SO(8,9)$ (this work was 
started before \cite{DH} and independently). Other instances of non
split forms of U-dualities occuring in pure supergravities in 4d 
the so-called 
magic triangle will be analysed in \cite{HJ} in order to check the precise 
control of chaos by symmetry in this more general situation. In many cases the 
simple rule of overextension of 3d U-duality is sufficient to analyse the 
chaotic or non chaotic behaviour of the flow. It is important to reach a more 
rigorous level of characterisation of the chaos and work is being done in this 
direction whereas theorems are available on the non-chaotic side \cite{AR}. 

Acknowledgements:
I am grateful to J. McKay and M. Henneaux for references and discussions.

\end{document}